\documentclass[letterpaper, 10 pt, conference]{ieeeconf}  

\IEEEoverridecommandlockouts                              

\let\labelindent\relax
\usepackage{graphicx} \graphicspath{ {figure/} }
\usepackage{amsmath}
\usepackage{amssymb}
\usepackage{nccmath}
\usepackage{multirow}
\usepackage{algorithmic}
\usepackage[ruled]{algorithm2e}
\usepackage{acronym}
\usepackage{xspace}
\usepackage{enumitem}
\usepackage{hyperref}
\usepackage{balance}
\usepackage{setspace}
\usepackage[skip=3pt]{subcaption}
\usepackage{caption}
\usepackage[misc]{ifsym}
\usepackage[dvipsnames]{xcolor}
\usepackage[capitalise]{cleveref}
\usepackage{overpic}

\makeatletter
\DeclareRobustCommand\onedot{\futurelet\@let@token\@onedot}
\def\@onedot{\ifx\@let@token.\else.\null\fi\xspace}
\def\eg{\emph{e.g}\onedot} 
\def\ie{\emph{i.e}\onedot} 
 
\def\etc{\emph{etc}\onedot}

\makeatother

\makeatletter
\def\BState{\State\hskip-\ALG@thistlm}
\makeatother

\makeatletter
\renewcommand{\paragraph}{%
  \@startsection{paragraph}{4}%
  {\z@}{0ex \@plus 0ex \@minus 0ex}{-1em}%
  {\hskip\parindent\normalfont\normalsize\bfseries}%
}
\makeatother

\crefname{algocf}{alg.}{algs.}
\Crefname{algocf}{Algorithm}{Algorithms}

\acrodef{vr}[VR]{Virtual Reality}
\acrodef{ar}[AR]{Augmented Reality}
\acrodef{vio}[VIO]{Visual Inertial Odometry}

\title{\LARGE \bf Congestion-aware Evacuation Routing using Augmented Reality Devices
}

\author{Zeyu Zhang$^{1}$\quad{}Hangxin Liu$^{1}$\quad{}Ziyuan Jiao$^{1}$\quad{}Yixin Zhu$^{1,2}$\quad{}Song-Chun Zhu$^{1,2}$\quad{}
\thanks{$^{1}$ UCLA Center for Vision, Cognition, Learning, and Autonomy (VCLA) at Statistics Dept. Emails:
{\tt\small \{zeyuzhang, hx.liu, zyjiao, yixin.zhu\}@ucla.edu}, \tt\small{sczhu@stat.ucla.edu}.}%
\thanks{$^{2}$ International Center for AI and Robot Autonomy (CARA)}%
\thanks{The work reported herein was supported by ONR MURI N00014-16-1-2007, DARPA XAI N66001-17-2-4029, and ONR N00014-19-1-2153.}%
}

\begin{document}

\maketitle
\thispagestyle{empty}
\pagestyle{empty}

\begin{abstract}
We present a congestion-aware routing solution for indoor evacuation, which produces real-time individual-customized evacuation routes among multiple destinations while keeping tracks of all evacuees' locations. A population density map, obtained on-the-fly by aggregating locations of evacuees from user-end \ac{ar} devices, is used to model the congestion distribution inside a building. To efficiently search the evacuation route among all destinations, a variant of A$^\star$ algorithm is devised to obtain the optimal solution in a \emph{single} pass. In a series of simulated studies, we show that the proposed algorithm is more computationally optimized compared to classic path planning algorithms; it generates a more time-efficient evacuation route for each individual that minimizes the overall congestion. A complete system using \ac{ar} devices is implemented for a pilot study in real-world environments, demonstrating the efficacy of the proposed approach.
\end{abstract}

\section{Introduction}

U.S. Federal Emergency Management Agency (FEMA) statistics~\cite{fema_stat} have revealed that 4.68 million in-building fires happened in a 10-year period of 2007 to 2016, along with 27,000 deaths and 139,925 fatal injuries. Although modern buildings often provide detailed 2D floor plans and clear emergency exit signs, tenants in general still take \emph{extra} time in wayfinding during emergencies~\cite{schwering2017wayfinding} due to fear, chaos, gridlock traffic, \etc, which jeopardizes their evacuation efficiency. In addition, such static information lacks real-time dynamics in terms of congestion, which further impedes the evacuation efficiency. Consequently, additional efforts must be spent on education~\cite{shiwaku2008proactive,kholshchevnikov2012study}, drills~\cite{ma2012experimental,xudong2009study}, or training in simulations~\cite{xi2014simulating,li2017earthquake} in advance to reduce casualties.

However, such preventative training could be costly in setup; it is also challenging to recreate and emulate the actual (dangerous) scenarios realistically. Such insufficiency urges the needs for an \emph{intuitive} system that can direct naive users with no prior experience or limited training to safely and effectively evacuate during actual emergencies.

In real-world settings, there are two main difficulties that prevent existing methods to form an effective means of evaluation. First, although the evacuation time has been identified as a vital factor~\cite{arbib2018applying} to improve evacuation efficiency, the majorities of the prior methods only consider the layout of the building (\eg, the distance between two key points, the width of a hallway), but not the congestion condition during the evacuation. Although the localization and navigation methods developed for indoor mobile robots~\cite{thrun2005probabilistic} could be adapted to guide evacuees~\cite{chittaro2008presenting,mulloni2012indoor}, and several localization methods using various devices or sensors can effectively provide \emph{individual}'s location, it remains challenging to take the \emph{overall} congestion into account in evacuation routing. Second, the process of perceiving, analyzing, and deciding the evacuation route needs to be in real-time for multiple users, requiring a time-efficient path planning algorithm that can handle a large number of queries.

\begin{figure}[t!]
    \centering
    \includegraphics[width=\linewidth]{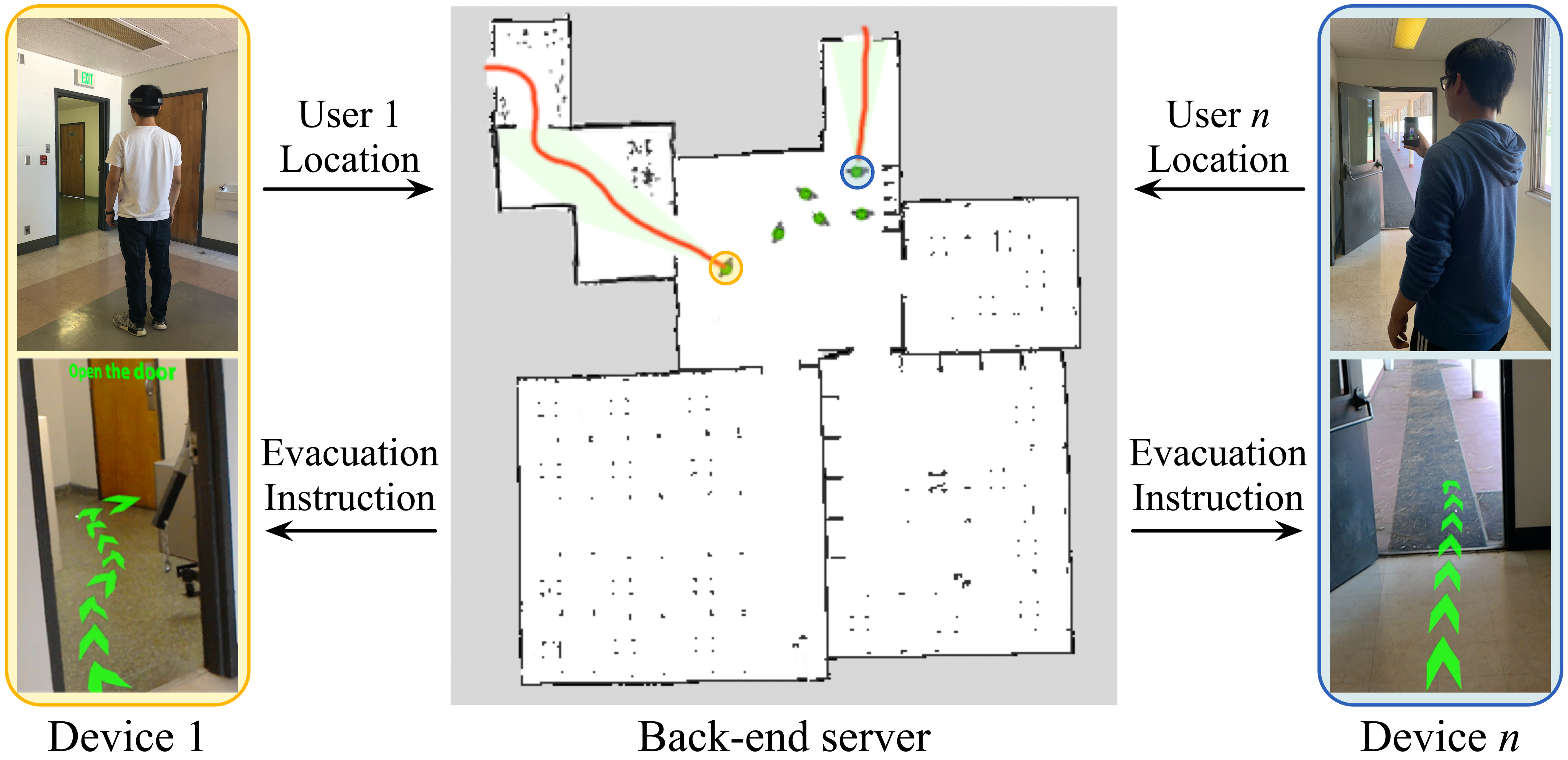}
    \caption{The architecture and data flow of the proposed system. By streaming the locations from the user-end devices (\eg, \ac{ar} headsets, cellphones), the remote server gathers population distribution of the environment in real-time. The centralized information is used to generate a time-efficient evacuation route and augment that to a user's view.}
    \label{fig:motive}
\end{figure}

To address these challenges, this paper leverages the recent advancement of cloud computing with the growing availability and throughput of the indoor wireless network. As illustrated in \cref{fig:motive}, by streaming the real-time localization from the edge devices, a population density map can be generated in a remote server by aggregating individual's locations; higher density indicates more congestion (see \cref{fig:simulation}). A congestion-aware evacuation routing algorithm is devised to provide the most efficient evacuation route by searching among \emph{all} possible destinations. A parallel asynchronous design of the proposed algorithm is further developed to support thousands of path planning queries in real-time. To validate the proposed method and system, a pilot study was conducted using a complete system implemented with the Microsoft HoloLens \ac{ar} platform.

\subsection{Related Work}

\textbf{Localization} of evacuees is a prerequisite in providing an efficient evacuation and modeling the congestion condition. In addition to a typical localization setup in robotics~\cite{thrun2005probabilistic}, RFID tags~\cite{chittaro2008presenting}, info points~\cite{mulloni2012indoor,mulloni2011handheld}, images~\cite{ahn2012indoor}, and the received WiFi~\cite{alnabhan2014insar,penmetcha2017smartresponse} or Bluetooth~\cite{shao2018marble} wireless signals are introduced to improve the localization accuracy under an emergency. However, they require some infrastructures deployed in advance and extra efforts at gathering localization information from individuals. \textbf{Multi-robot localization} methods can obtain each robot's location via a peer-to-peer broadcasting~\cite{roumeliotis2002distributed,nerurkar2009distributed}, but they are usually slow in convergence, and the communication network between robots could be unreliable in practice. To address such deficiencies, the present work localize edge users using a SLAM-based method and a cloud computing platform to gather individuals' locations in a centralized manner.

Typical \textbf{path planning} algorithms (\eg, A$^\star$, RRT$^\star$) can be used for evacuation routing by minimizing evacuation distance~\cite{kim2008vision,liu2016xyz,gerstweiler2018dargs} or maximizing time efficiency~\cite{zu2017distributed,wong2017optimized,liu2018path} to properly handle the planning requests from a large number of agents~\cite{velagapudi2010decentralized,liu2014dynamic}. However, one of the key assumptions is that the congestion condition is fixed and provided upon planning. To take congestion into account, the field of \textbf{multi-agent path finding} has developed various approaches to find collision-free paths for multiple robots/agents~\cite{sharon2015conflict,ma2017multi}. However, such approaches are still brittle in large-scale due to limited capability of handling dynamic changes of the congestion condition, possibly because adaptation to the changes of the congestion condition in real-time is a non-trivial problem, requiring additional engineering efforts. To handle hundreds of requests in real-time with dynamic changes of the congestion condition, the system presented in this paper jointly optimizes both distance and time by considering the overall congestion on-the-fly and developed a paralleled planning scheme.

\textbf{\ac{ar}} technologies have received a considerable amount of attention in various human-robot interaction settings~\cite{zaeh2006interactive,zolotas2018head,liu2018interactive}. It affords an intuitive interaction to a naive user by overlaying rich \emph{virtual} visual aids on top of the observed \emph{physical} environment. In literature, several evacuation systems using \ac{ar} technologies have been proposed, though limited, to provide indoor evacuation assistance. Typical \ac{ar} devices, such as \ac{ar} headsets~\cite{rehman2015augmented,sanchez2016participatory,gerstweiler2018dargs}, are equipped with depth cameras and IMUs, capable of providing a precise indoor localization. Due to these advantages, this paper implements and validates the system using the state-of-the-art \ac{ar} headset, Microsoft HoloLens. 

\subsection{Contribution}

This paper makes the following three contributions:
\begin{enumerate}[leftmargin=*,noitemsep,nolistsep]
    \item We propose and implement a real-time population density map that models the \emph{congestion} during the evacuation. The population distribution is aggregated from decentralized user-end devices to a centralized cloud server.
    \item We propose and implement an efficient congestion-aware evacuation routing algorithm that simultaneously searches among \emph{all} possible destinations in a \emph{single} pass, providing the most time-efficient route by utilizing the population density map. A parallelized asynchronous solution of the algorithm is also devised to support thousands of path planning queries in real-time. 
    \item We prototype a complete evacuation system using the Microsoft HoloLens as the user-end device. A pilot study has been conducted to validate the viability and efficacy of the proposed algorithm and system.
\end{enumerate}

\subsection{Overview}

The remainder of the paper is organized as follows. \cref{sec:density_map} introduces the construction of the population density map. The congestion-aware evacuation routing algorithm is described in \cref{sec:plan}. \cref{sec:exp} evaluates the proposed system in both simulation and real-world scenarios. \cref{sec:discussion} concludes the paper with discussions.

\begin{figure}[t!]
    \centering
    \includegraphics[width=\linewidth]{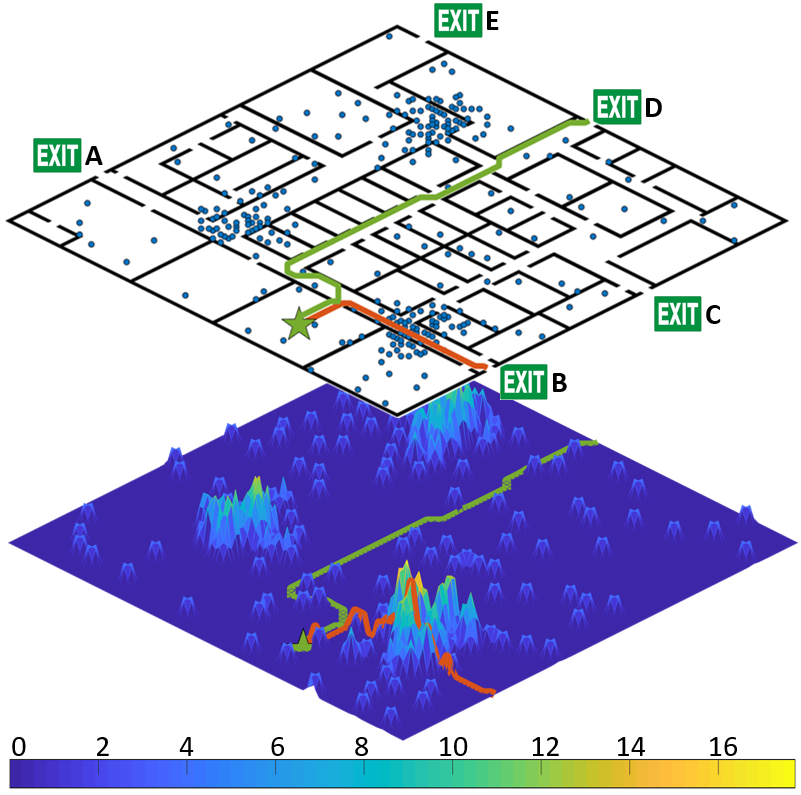}
    \caption{Evacuation routing with a density map. (Top) The floorplan and the distribution of evacuees (blue dots). (Bottom) The population density map indicates the magnitude of congestion. If a human agent (green star) followed the route generated by a naive planner in terms of the shorted path (in red) to Exit B, s/he would compete with other agents. Instead, the proposed system suggests a further, but more time-efficient evacuation route (in green) toward Exit D.}
    \label{fig:simulation}
\end{figure}

\section{Population Density Map}\label{sec:density_map}

Without the congestion information, individual evacuees would only be able to choose a route by minimizing the distance, incapable of considering the actual time needed. Hence, an effective choice of evacuation route during emergencies should take the congestion into account. This section describes the proposed method that models the congestion as a population density map aggregated from user-end devices.

\subsection{Localization}

Given a 2D floorplan, various SLAM methods could be used for localization. Since devising a SLAM method is out of the scope of this paper, an off-the-shelf solution based on ORB-SLAM~\cite{mur2017orb} is adopted and briefly described below. 

The extracted ORB features~\cite{rublee2011orb} are rotation-invariant and noise-resistant. The features yield a robust representation to camera condition changes (\eg, auto-gain, auto-exposure, illumination changes), which is an important property in emergency settings due to the varied lighting and rapid motions in evacuations. A feature matching process is performed to produce a set of monocular keypoints. The DBoW2 algorithm~\cite{galvez2012bags} further matches these key points to a keyframe of the environment by searching through the pre-built covisibility graph~\cite{strasdat2011double}. A motion-only bundle adjustment~\cite{mur2017orb} uses the best-matched keyframe to determine and optimize the camera pose, thus providing an evacuee's location. This process can be performed on typical \ac{ar} headsets with built-in cameras (\eg, HoloLens) or smartphones~\cite{mulloni2012indoor,mulloni2011handheld,ahn2012indoor}.

\subsection{Population Distribution and Density Map}

The population distribution is directly related to congestion during emergencies. After performing the localization on user-end devices (\eg, \ac{ar} headsets, mobile phones), the population distribution in the environment can be obtained by aggregating all users' locations to the cloud server through the wireless local area networks.

A density map can be further computed based on the population distribution to describe the congestion condition in the environment. The density map is a two-dimensional grid-like map built on top of the floorplan. \cref{fig:simulation} shows an example of the density map with its corresponding population distribution; the color bar indicates the magnitude of congestion $\rho$ at location $(x, y)$:
\begin{equation}
    \rho_{x,y} = \sum_{(x_i, y_i) \in \mathcal{C}_{x, y}} \frac{\gamma N_{x_i, y_i}}{\sqrt{2\pi}} e^{-\frac{(x_i - x)^2 + (y_i - y)^2}{2}},
\end{equation}
where $\mathcal{C}_{x, y}$ is a set of cells which forms a patch centered at coordinate $(x, y)$, $N_{\cdot, \cdot}$ the number of agents, and $\gamma$ the congestion coefficient, describing the influence radius of every agent. The value of $\gamma$ depends on the resolution of the grid map; we set $\gamma=5$ in our experiment. The path planning algorithm proposed in the subsequent section utilizes this density map to estimate the congestion while planning an evacuation route that better balances the distance and time.

\section{Evacuation Routing with Density Map}\label{sec:plan}

Consider the scenario shown in \cref{fig:simulation}; if a naive planner were used, the agent (green star) would follow the red path to Exit B as it is the shortest. However, though further, it may be more favorable to the agent to follow the green path leading to Exit D with less congestion. This example illustrates the necessity of taking the density map into account in evacuation routing. This section describes the proposed \emph{congestion-aware evacuation routing} algorithm that accounts for the density of other evacuees in a route and performs a centralized traffic control to avoid congestion and reduce competition during the evacuation in an emergency.

At a high level, this algorithm is a variant of the classic A$^\star$ algorithm, which is designed for pathfinding in multi-destination scenarios. It is modified (i) to search and select the most time-efficient path among \emph{all} candidate destinations in a \emph{single} pass, instead of searching individual destination with multiple passes as classic A$^{\star}$ does, and (ii) to avoid explore all of the possible destinations due to the bounded cost: during each iteration, the algorithm only explores the grid that has the smallest F-score along the direction of the exits/destinations; see details of the algorithm in below.

\paragraph*{Data Structure}

In practice, the explored space is much smaller compared to the exhausted Dijkstra search or A$^\star$ search with multiple passes. The algorithm is outlined in \cref{alg:md_astar}; it is performed based on a data structure \emph{Node}. Each Node $p$ contains:
\begin{enumerate}[leftmargin=*,noitemsep,nolistsep]
    \item Position ($p.pos$) stores the $(x, y)$ coordinates of the node in the grid map.
    \item Parent ($p.parent$) records the $(x, y)$ coordinates from which $p.pos$ comes; $p.parent = null$ indicates current node is the start node. 
    \item Destination ($p.dst$) keeps the coordinates of the destination.
    \item Total cost ($p.c$) represents the total cost from the start position to $p.pos$.
    \item Heuristic score ($p.h$) indicates how far from $p.pos$ to $p.dst$ heuristically.
    \item F-score ($p.f$) depicts the total estimated cost from the start node to $p.dst$. Specifically, we have $p.f = p.c + p.h$.
\end{enumerate}

\begin{algorithm}[t!]
    \caption{Congestion-aware Evacuation Routing} 
    \label{alg:md_astar}
    \LinesNumbered
    
    \SetKwInOut{KIN}{Input}
    \SetKwInOut{KOUT}{Output}
    
    \KIN{Grid map: $G_{grid}$\\
        Density map: $G_{density}$\\
        Start point: $src$\\
        Destinations: $DList = \{dst_1, dst_2, \ldots, dst_n\}$
    }
    \KOUT{A selected path from $src$ to one of the destinations $dst_k$}
    
    Initialize the open list: $OpenList = \emptyset$
    
    Initialize the closed list: $ClosedList = \emptyset$
    
    Insert Node($src$, $null$) to $OpenList$

    \While{$OpenList$ is not empty}{
        // Pop node with the least f-score
        
        $p$ $\leftarrow$ $OpenList.pop()$
        
        // Reach the best destination
        
        \If{$p == dst$}{
            \Return    $TracePath(dst)$
        }
        
        $S$ $\leftarrow$ $FindSuccessors(p, G_{grid})$
        
        \ForEach{$ s \in S$}{
            \ForEach{$dst \in DList$}{
                $s.dst$ $\leftarrow$ $dst$
                
                $s.g$ $\leftarrow$ $p.g$ + $Cost(p, s, G_{density})$
                
                $s.h$ $\leftarrow$ $Heuristic (s, dst)$
                
                $s.f$ $\leftarrow$ $s.g$ + $s.h$
                
                \If{$Validate(s, OpenList, ClosedList)$}{
                    Insert $s$ to $OpenList$ 
                }
            }
        }
        
        Insert $p$ to $ClosedList$
    }
\end{algorithm}

\paragraph*{Implementation details}

The algorithm maintains two lists; \ie, the $OpenList$ and the $ClosedList$. The $OpenList$ stores nodes waited to be explored, and the $ClosedList$ records nodes already visited. Each is implemented by a heap-based priority queue, which stores all nodes in order, and is further augmented with a hash table to look up one of its elements in constant time. Both $OpenList$ and $ClosedList$ support two operations: (i) \emph{pop()} returns a node with the least F-score and removes the node from the list, and (ii) \emph{insert(p)} inserts the node $p$ into the list. See \cref{alg:md_astar} for an outline of the algorithm. Below, we detail some important functions.

\begin{itemize}[leftmargin=*,noitemsep,nolistsep]
    \item $Node(a, b)$ creates a new $Node$, whose $pos$ field is $a$ and $parent$ field is $b$.
    \item $TracePath(dst)$ retraces the path from the start position to $dst.pos$ by recursively visiting the $parent$ field.
    \item \emph{FindSuccessors($p$, $G_{grid}$)} looks up the grid map $G_{grid}$ and returns a list of nodes whose position is adjacent to $p$ in the $G_{grid}$. By default, the $parent$ field is set to $p$.
    \item The heuristic function \emph{Heuristic($s$, $dst$)} is defined as the euclidean distance between $s.pos$ and $dst.pos$.
    \item \emph{Validate(s, OpenList, ClosedList)} determines whether the node $s$ should be inserted into the $OpenList$. The node $s$ would be added to $OpenList$ if one of the following conditions is satisfied: (i) $s$ is not in the $OpenList$; (ii) a node in the $OpenList$ whose position is as same as $s$ but has a higher heuristic score $f$; and (iii) a node in the $ClosedList$ whose position is as same as $s$ but has a higher total cost $c$. The key ideas behind these conditions are: (i) if the node has not been explored, add it to the $OpenList$; (ii) the node $s$ may have a better solution than the one at the same position in the $OpenList$ to be explored; and (iii) there is a better solution found to reach position $s.pos$, therefore re-explore the position $s.pos$.
    \item \emph{Cost($p$, $s$, $G_{density}$)} calculates the cost of travelling from $p.pos$ to $s.pos$. The cost $c$ is calculated by
\begin{equation}
    \label{eqn:cost_function}
    c = \beta \cdot \text{dist}(s.pos, p.pos) + (1 - \beta) \cdot G_{density}[s.pos],
\end{equation}
    where $\text{dist}(s.pos, p.pos)$ is the euclidean distance, $G_{density}[s.pos]$ looks up the density map and returns the congestion estimation at position $s.pos$, $\beta$ is a coefficient that balances between distance and congestion estimation.
\end{itemize}

\paragraph*{Paralleled Asynchronous Planning}

\begin{figure}[b!]
    \centering
    \includegraphics[width=\linewidth]{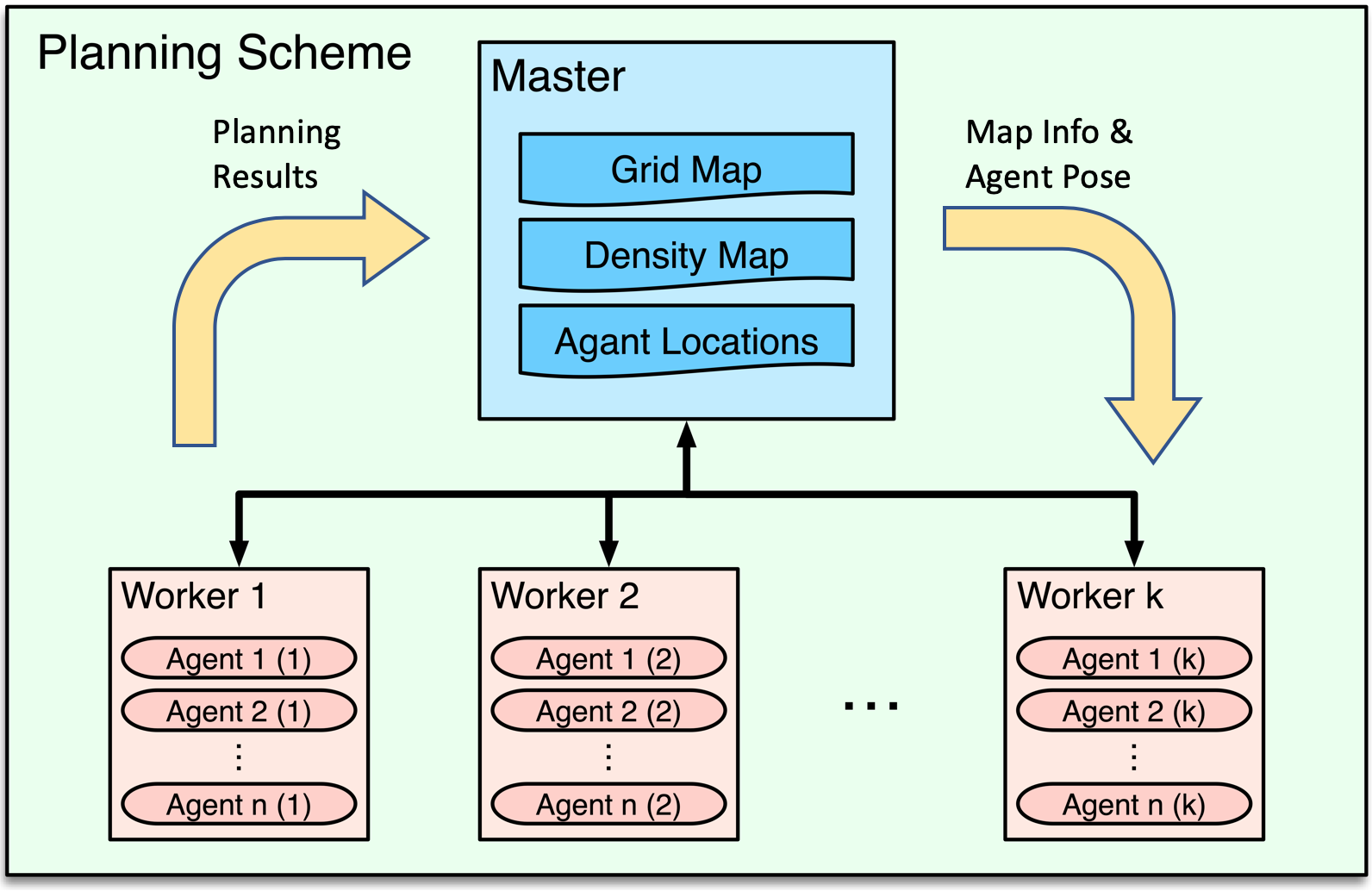}
    \caption{Illustration of the parallelized asynchronous planning scheme. Both \emph{Workers} and \emph{Master} run in separated processes on a same multi-core machine.}
    \label{fig:planning_arch}
\end{figure}

\begin{figure}[b!]
    \centering
    \begin{subfigure}[b]{0.5\linewidth}
        \includegraphics[width=\linewidth]{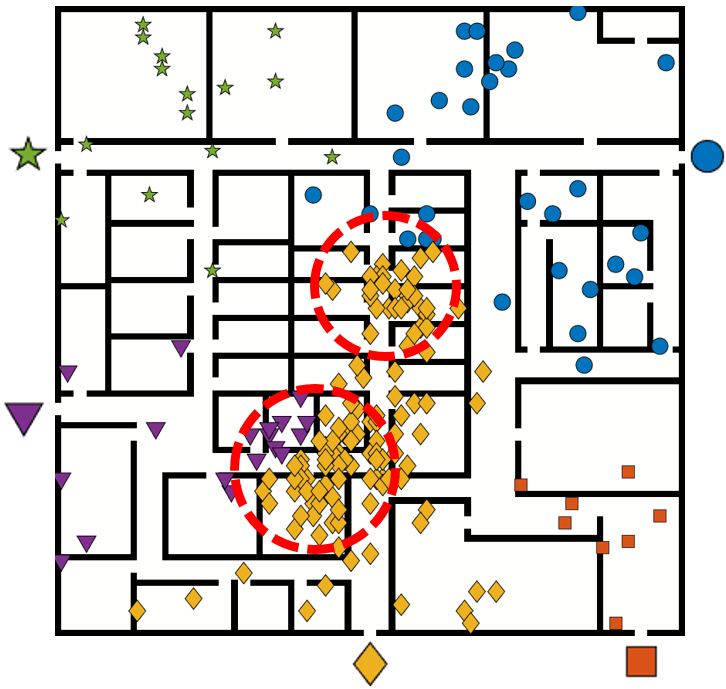}
        \caption{}
    \end{subfigure}%
    \begin{subfigure}[b]{0.5\linewidth}
        \includegraphics[width=\linewidth]{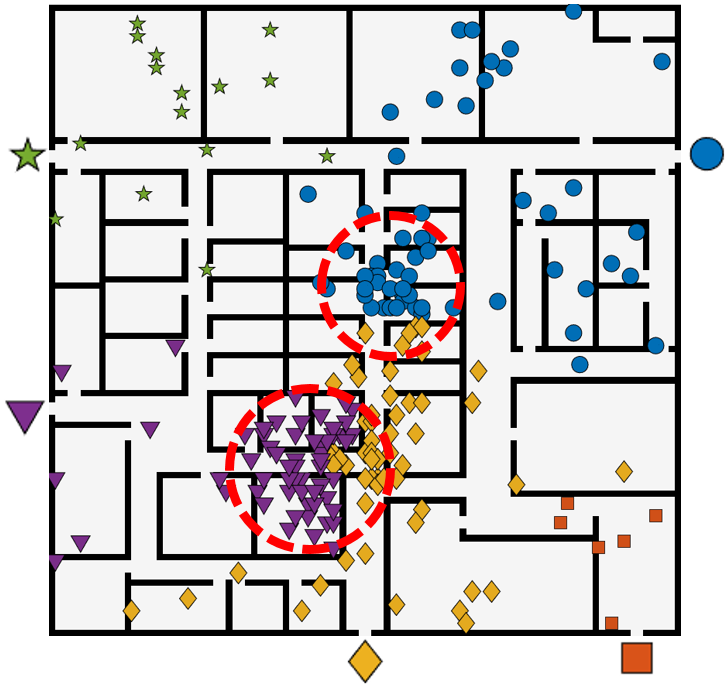}
        \caption{}
    \end{subfigure}
    \caption{Qualitative comparison of routing results. The agents' marker types are assigned based on the exits they are directed to. (a) Assigning the agents to their closest exit leads to congestion and competition. (b) The proposed method directs the agents in the dashed circles to another exit, though further, to avoid congestion; thus is more time-efficient.}
    \label{fig:planning_result}
\end{figure}
 
A paralleled asynchronous planning scheme is introduced to more efficiently schedule the planning jobs requested by user-end devices in a remote cloud server. \cref{fig:planning_arch} shows the high-level design. The users' job requests are evenly assigned to multiple \emph{workers} (\eg, CPU cores). Every \emph{worker} maintains a list of path planning requests and processes them in turns. A \emph{worker} periodically pushes the result back to the \emph{master}, who maintains user information, updates the density map, and provides the latest population density to the user-end devices. Such a design can significantly improve the planning efficiency for a large number of users' planning requests by fully exploiting the computation power of remote servers.

\section{Experiments}\label{sec:exp}

We first evaluated the proposed method in a simulated indoor environment with crowds simulation to validate (i) the necessity of the introduced density map for modeling the congestion in terms of egress time, and (ii) the computational efficiency of the proposed planning algorithm comparing to classical planning algorithms. 

The method is implemented in a proof-of-concept system consisting of Microsoft HoloLens as the user-end device and a regular blade server as the remote server. Additional experiments are conducted in the physical environment to qualitatively show the visual aids rendered by the system and quantitatively evaluate the system through a pilot study. 

\subsection{Simulation Setup}

A simulated environment (see \cref{fig:simulation}) as a 2D map is constructed. The environment has a dimension of 100m$\times$100m, which is discretized into 100$\times$100 grids. It simulates a floorplan of a typical school building that consists of classrooms, offices, study rooms, auditoriums, \etc. There are five exits (EXIT A-E) in the environment.

The microscopic behaviors of the evacuees, such as the speed of their movement affected by others, are also crucial for a proper simulation in order to more realistically estimate the overall evacuation efficiency. Given the population distribution, the evacuees' movement is simulated by their gait length using the SLIP model~\cite{blickhan1989spring,geyer2006compliant}, which has been widely used to simulate bipedal locomotion; the maximum allowable human gait length is computed to bound an evacuee's movement and velocity.

\subsection{Evacuation Efficiency}

\begin{figure}[t!]
    \centering
    \includegraphics[width=\linewidth,trim={0.65cm 0cm 1.3cm 0.0cm},clip]{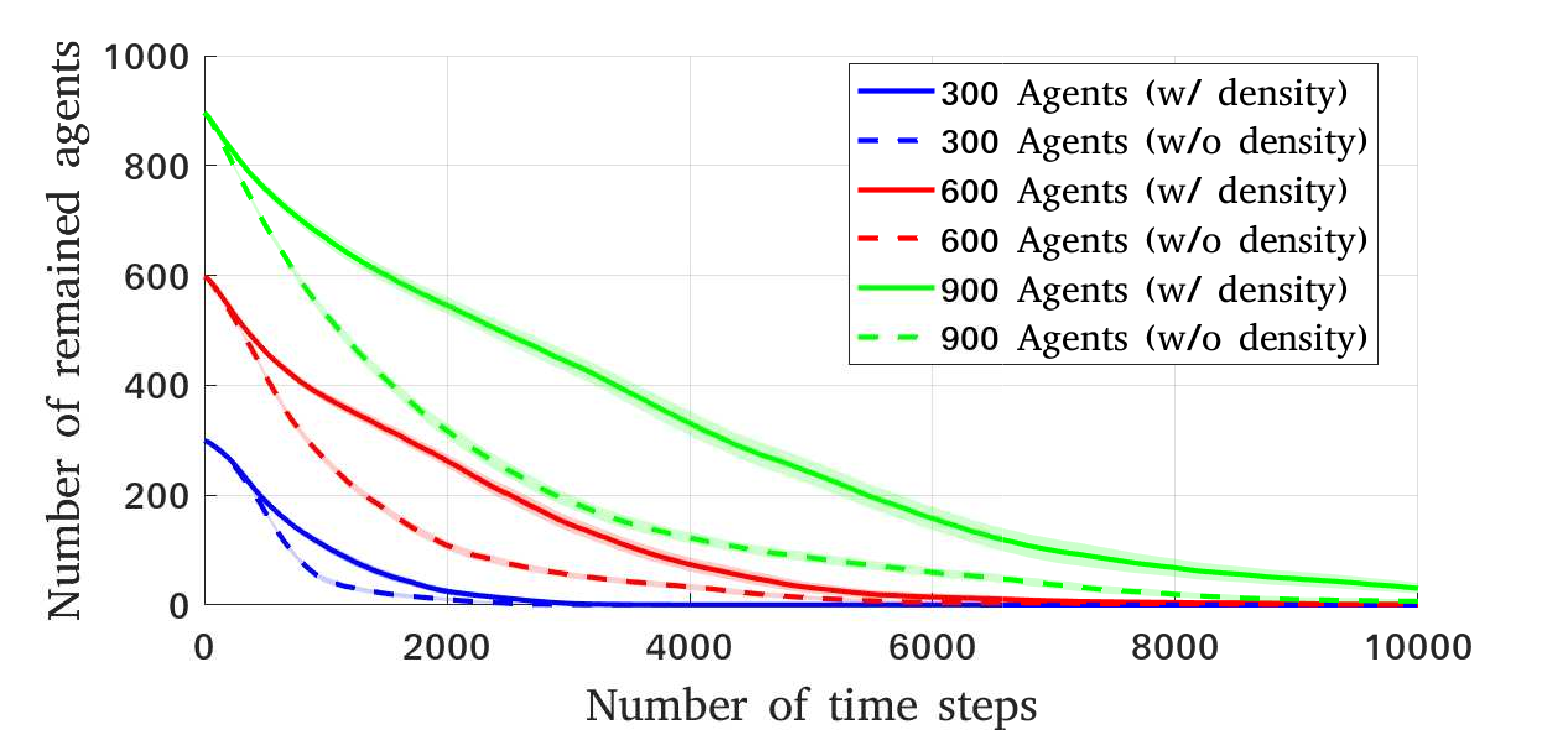}
    \caption{Quantitative comparison of the average number of remained agents at each time step. The shaded color strips indicate the 98\% confidence interval over 100 trails.}
    \label{fig:simu_n_agents}
\end{figure}

\begin{figure}[t!]
    \centering
    \begin{subfigure}[b]{\linewidth}
        \includegraphics[width=\linewidth]{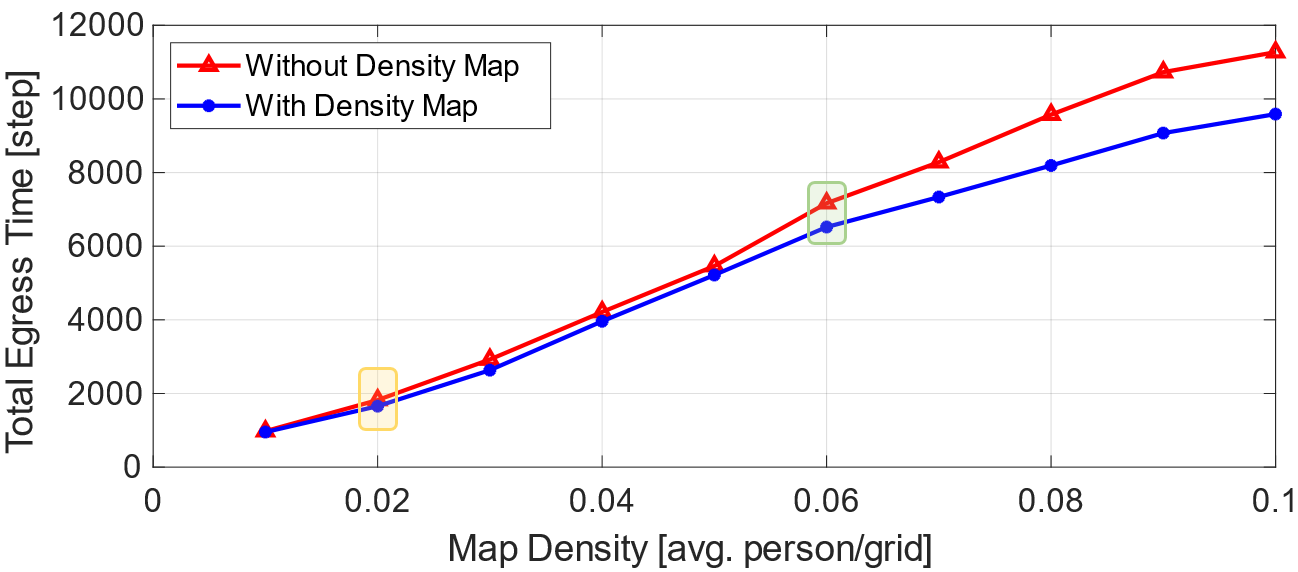}
        \caption{}
    \end{subfigure}%
    \\
    \begin{subfigure}[b]{0.5\linewidth}
        \includegraphics[width=\linewidth]{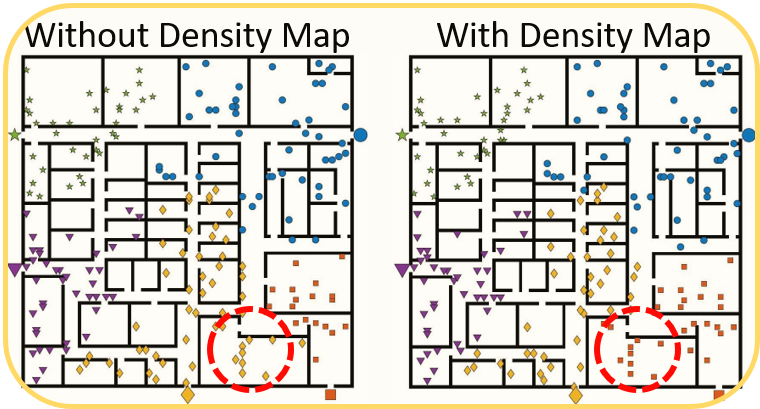}
        \caption{}
    \end{subfigure}%
    \begin{subfigure}[b]{0.5\linewidth}
        \includegraphics[width=\linewidth]{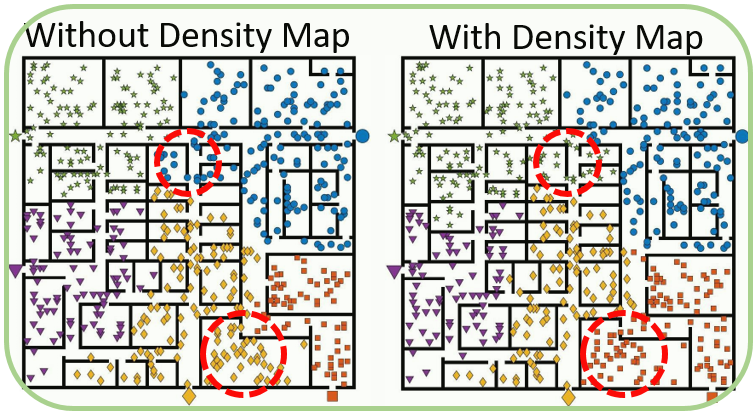}
        \caption{}
    \end{subfigure}
    \caption{(a) Average total egress time vs. map density---average number of people per grid---of the scenario. Each point represents an average of 250 runs. (b)(c)The points in yellow and green shaded box denote examples of planning results for 0.02 and 0.06 map density, respectively.}
    \label{fig:simu_ave_time}
\end{figure}

Using the simulation environment, the effectiveness of the congestion modeling is evaluated by the total egress time--the time for all agents to evacuate from the environment. \cref{fig:planning_result} highlights the differences of the routing results between simply finding the closest exits in terms of the distance and the proposed method that directs agents to avoid congestion by considering both distance and time. \cref{fig:simu_n_agents,fig:simu_ave_time} further quantitatively showcase the efficacy of the proposed method; it yields a faster evacuation at every step and the final egress time, especially when the population density is high.

\subsection{Computation Efficiency}

\begin{figure}[t!]
    \centering
    \includegraphics[width=\linewidth,trim={1.5cm 0.1cm 1.6cm 1.5cm},clip]{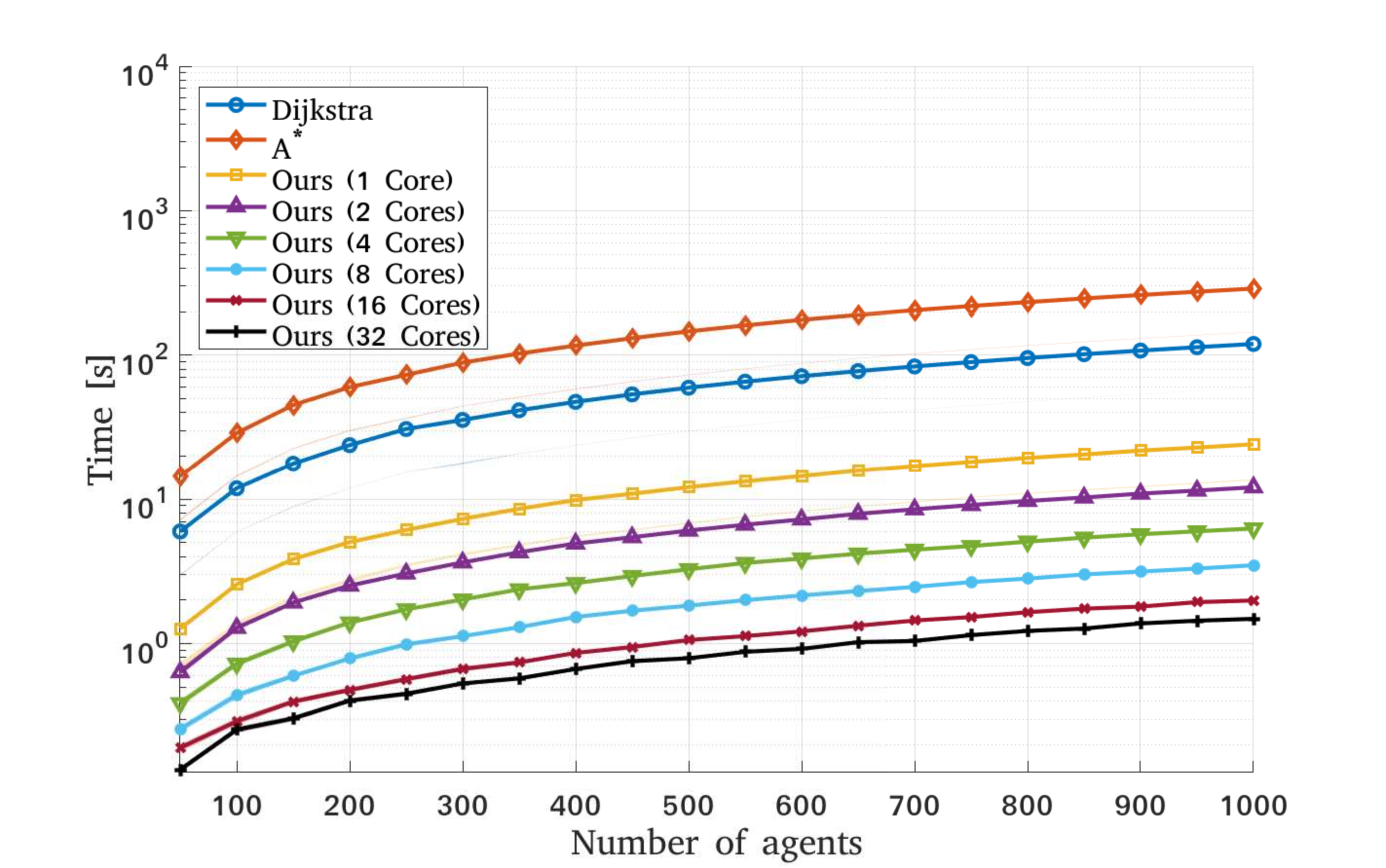}
    \caption{The proposed method is more time-efficient compared to Dijkstra and repeated A$^\star$ path planning algorithm in a single-core setting. It can be further paralleled to multi-core computing setup with a significant performance gain, achieving a real-time performance (around 1 second) to serve 1,000 users using 32 cores.}
    \label{fig:calctime}
\end{figure}

\cref{fig:calctime} shows the average run time (in log-scale) of 50 trails to generate the evacuation routing for up to 1,000 agents, wherein we compare the proposed method with A$^\star$ algorithm and Dijkstra algorithms. Although the proposed algorithm runs significantly faster than classic approaches using a single core, it still takes more than 10 seconds to serve 1,000 users. This insufficiency is the core motivation to introduce the paralleled asynchronous planning scheme to take advantage of modern multi-core computers. The proposed algorithm can be easily paralleled and distributed into different cores; the computation time of using a different number of cores (2-32) is shown in \cref{fig:calctime}. In our experiments, the proposed method can achieve real-time performance to serve 1,000 users with 32 cores.

\subsection{System Prototype}

The prototype system adopts the state-of-the-art \ac{ar} head-mount display, Microsoft HoloLens, as the user-end device. Compared to other available \ac{ar} headsets, HoloLens is the first untethered \ac{ar} head-mounted display that allows the user to move freely in the space without being constrained by cable connections, which is particularly crucial for evacuation applications. Integrated with 32-bit Intel Atom processors, HoloLens equips with an IMU and multiple spatial-mapping cameras to perform low-level computation onboard. Using Microsoft's Holographic Processing Unit, the users can realistically view the augmented contents. The holograms displayed on its screen are created by Unity3D, via which various visual effects can be rendered.

A back-end server is deployed to handle computationally intense jobs (\eg, density map construction, path planning). The system in the server hosts a task scheduler, which coordinates the computation sources and assigns the new-coming user to different worker threads (see \cref{fig:planning_arch}). Once the remote server receives users' locations, an evacuation path for every user is generated (see \cref{sec:plan}) and communicated back to the corresponding user-end device.

\begin{figure}[t!]
    \centering
    \begin{subfigure}[b]{\linewidth}
        \includegraphics[width=\linewidth]{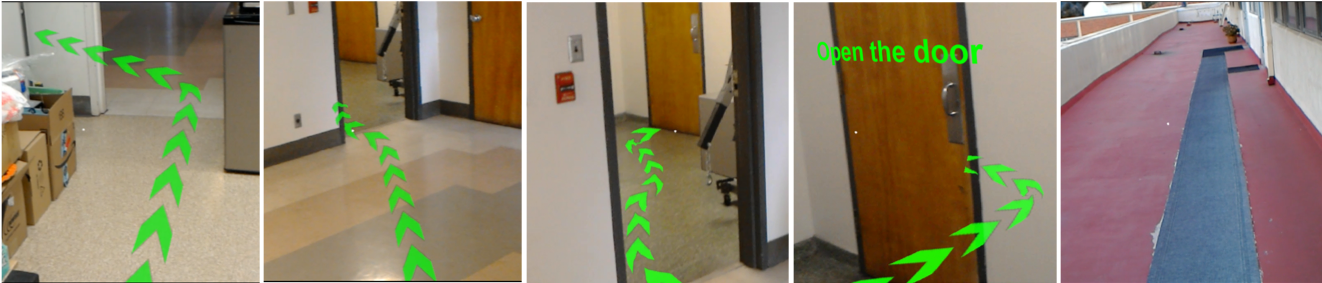}
        \caption{}
    \end{subfigure}%
    \\
    \begin{subfigure}[b]{\linewidth}
        \includegraphics[width=\linewidth]{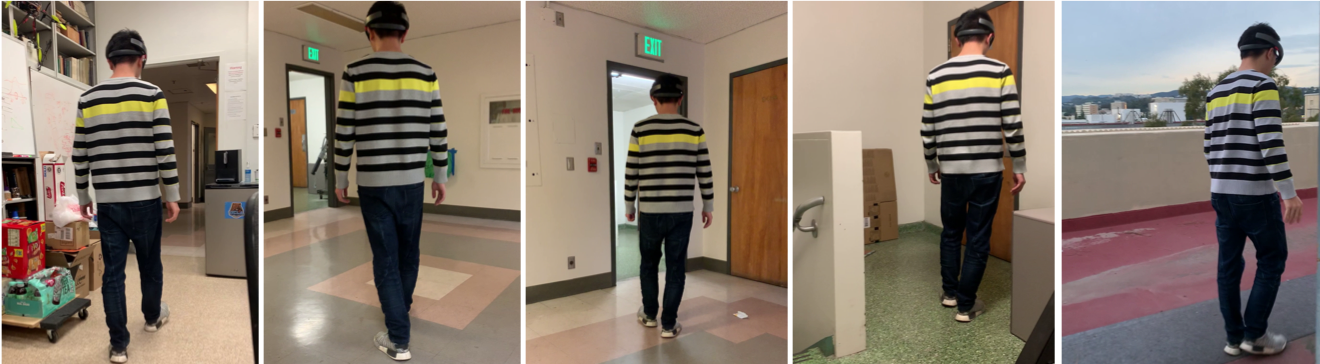}
        \caption{}
    \end{subfigure}%
    \\
    \begin{subfigure}[b]{\linewidth}
        \includegraphics[width=\linewidth]{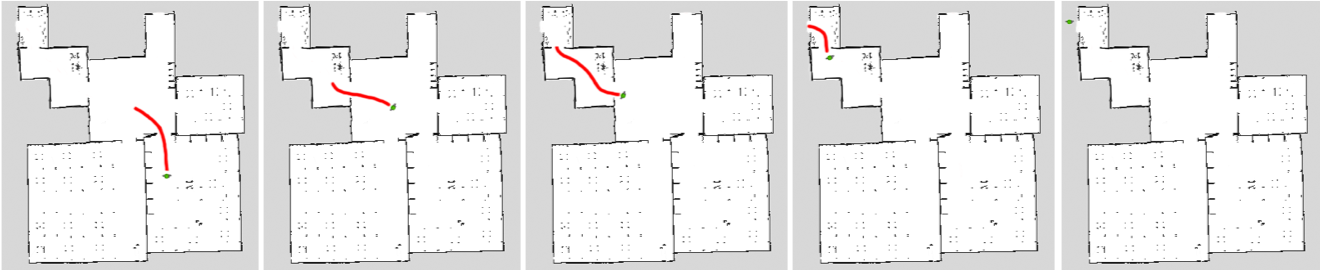}
        \caption{}
    \end{subfigure}%
    \caption{An example of the evacuation process using an AR headset in real world. (a) Planned path from the user's egocentric view using a Hololens \ac{ar} headset. (b) Third-person view. (c) Localization in the physical world.}
    \label{fig:exp_localization}
\end{figure}

As a proof-of-concept prototype, the system incorporates two types of visual guidance: (i) the evacuation path to direct a user to the most-efficient exit, which is updated on-the-fly as the user's location changes, and (ii) the visual symbols instruct the users to interact with the environment during the evacuation; for instance, as shown in the top row of \cref{fig:exp_localization}, a 3D text, ``Open the door,'' is augmented to the door to guide users' critical actions, further improving a user's reliance.

\subsection{Human Study}

We conducted a pilot study by recruiting 16 participants to evaluate the effectiveness of the \ac{ar} evacuation system in a between-subject setting ($N=8$ for each group). Half of the participants were in the baseline group and provided only a 2D physical floorplan; they were asked to find a path to evacuate in the physical world without any additional information. The other half of the participants were in the \ac{ar} group, using the \ac{ar} headset HoloLens with the proposed evacuation system installed. Each subject has no familiarization with the physical environments. The subjects in \ac{ar} group have no prior experience using \ac{ar} devices.

\begin{figure}[t!]
    \centering
    \includegraphics[width=\linewidth,trim={0.0cm 0.0cm 1.5cm 0.3cm},clip]{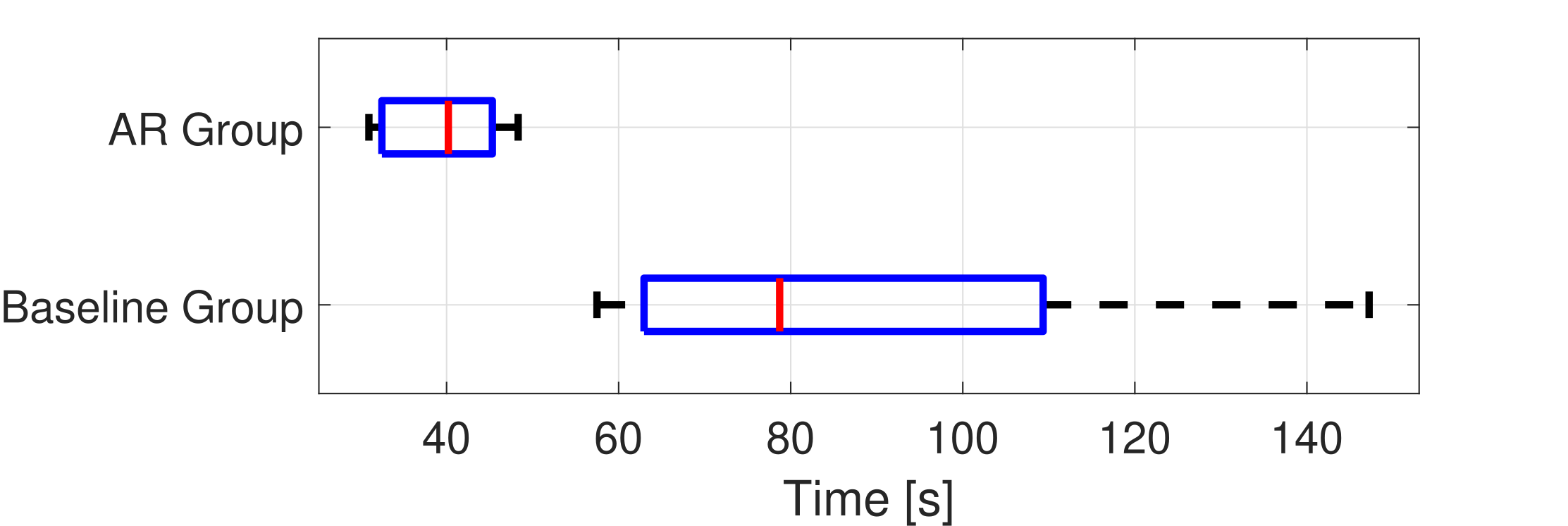}
    \caption{Box plot for all participants' escape time in two groups. Subjects in \ac{ar} group take a significantly less time.}
    \label{fig:_result}
\end{figure}

\cref{fig:exp_localization} depicts examples of the evacuation process, where the planned route from the user's egocentric view is shown in the top row, a third-person view is shown in the middle row, and a real-world localization is shown in the bottom row. \cref{fig:_result} compares the results between the two groups. The difference of escape time is statistically significant; $t(15)=1.0$, $p=0.0056$. Participants in \ac{ar} group takes a significantly less time (median: 40 seconds) to evacuate. In contrast, the baseline group requires much more time (median: 80 seconds) with a larger variance in terms of the evacuation time. This result indicates the efficacy of using the \ac{ar} evacuation system in a real-world evacuation scenario.

\section{Conclusion and Discussion}\label{sec:discussion}

This paper proposes a congestion-aware evacuation routing solution using augmented reality. To alleviate the overall congestion during emergencies, the proposed system simultaneously tracks of all evacuees' locations in real-time to provide a time-efficient evacuation path for each individual on-the-fly. The proposed method adopts the idea of edge computing that leverages the computation power on both user-end devices and the remote server. To further accelerate the proposed method, a parallel asynchronous planning scheme is devised to fulfill the demands of thousands of planning queries from evacuees. The simulated experiments demonstrated the effectiveness and efficiency of the proposed method by achieving real-time evacuation routing for 1,000 users. The proposed method is also implemented on a physical system using a Microsoft HoloLens and a remote server. A pilot study has been conducted to demonstrate the efficacy of the proposed method further.

Below, we discuss four related topics in greater depth.

\paragraph*{Alternating between different routes}
The proposed method updates evacuation routes as the density map updates. When the density map changes dramatically, \eg, a large group of users suddenly connect to the system, the system may suggest a completely different route and even alternate back-and-forth. In this case, a possible solution would be imposing extra constraints (\eg, a discount factor) when re-routing an evacuee to another exit. Alternatively, the density map only updates gradually as evacuees move around; thus, minor disturbance of the system is unlikely to cause such an alternation.

\paragraph*{Integrating predictive modules in routing}
In future work, it is possible to integrate a predictive model (\eg, Hidden Markov Model) for evacuees' movements, therefore achieving a better estimation of the density map. Such a predictive module could improve evacuation efficiency by resolving potential congestion in advance.

\paragraph*{Network accessibility}
The presented system currently relies heavily on the wireless local area network (\eg, WiFi) due to the required communications between the back-end server and HoloLens front ends. Under the situation where such infrastructure is not available, alternative forms of infrastructure (\eg, the wireless ad-hoc network, BlueTooth) could be easily adapted to support the communication.

\paragraph*{Adapting to other robot systems}
The proposed method and the paralleled asynchronous planning architecture could be further adapted to other large-scale multi-agent robot systems, which can leverage the power of distributed edge computing and a centralized back-end server.

\setstretch{1.2}

\end{document}